\documentclass[aps,prb,twocolumn,superscriptaddress]{revtex4}

\usepackage{amsmath}
\usepackage{graphicx}
\usepackage{amssymb}

\makeatletter

\begin{document}

\title{Uncovering the Dominant Scatterer in Graphene Sheets on SiO$_2$}

\author{Jyoti Katoch}

\affiliation{Department  of Physics,  University  of Central  Florida,
  Orlando, FL 32816-2385, USA}

\affiliation{Nanoscience Technology Center, University of Central
  Florida, Orlando, FL 32816-2385, USA}

\author{J.-H. Chen}

\affiliation{Department of Physics, University of Maryland, College
  Park, MD 20742, USA}

\author{Ryuichi Tsuchikawa}

\affiliation{Department  of Physics,  University  of Central  Florida,
  Orlando, FL 32816-2385, USA}

\affiliation{Nanoscience Technology Center, University of Central
  Florida, Orlando, FL 32816-2385, USA}

\author{C. W. Smith}

\affiliation{Department  of Physics,  University  of Central  Florida,
  Orlando, FL 32816-2385, USA}

\affiliation{Nanoscience Technology Center, University of Central
  Florida, Orlando, FL 32816-2385, USA}

\author{E. R. Mucciolo}

\affiliation{Department of Physics, University of Central Florida,
  Orlando, FL 32816-2385, USA}

\author{Masa Ishigami}

\affiliation{Department  of Physics,  University  of Central  Florida,
  Orlando, FL 32816-2385, USA}

\affiliation{Nanoscience Technology Center, University of Central
  Florida, Orlando, FL 32816-2385, USA}

\date{\today{}}

\begin{abstract}
We have measured the impact of atomic hydrogen adsorption on the
electronic transport properties of graphene sheets as a function of
hydrogen coverage and initial, pre-hydrogenation field-effect
mobility. Our results are compatible with hydrogen adsorbates inducing
intervalley mixing by exerting a short-range scattering potential. The
saturation coverages for different devices are found to be
proportional to their initial mobility, indicating that the number of
native scatterers is proportional to the saturation coverage of
hydrogen. By extrapolating this proportionality, we show that the
field-effect mobility can reach $1.5 \times 10^4$ cm$^2$/V sec in the
absence of the hydrogen-adsorbing sites. This affinity to hydrogen is
the signature of the most dominant type of native scatterers in
graphene-based field-effect transistors on SiO$_2$.
\end{abstract}

\maketitle
\bibliographystyle{apsrev}


Freely suspended graphene sheets display high field-effect mobility,
reaching $2 \times 10^5$ cm$^2$/V sec.\cite{bolotinss,bolotinprl} High
mobilities allows for a wider utilization of graphene sheets in
testing relativistic quantum mechanics, exploring two-dimensional
physics, and creating new electronic, optoelectronic, and spintronic
device
technologies.\cite{antonioreview,geimupdatescience,geimnovoselov} Yet,
suspended graphene sheets are fragile and impractical for most
experiments and applications. Substrate-bound graphene sheets are
easier to handle but possess low carrier mobilities, which can even
vary by an order of magnitude from sample to sample. Poor and
unpredictable transport properties reduce the utility of
substrate-bound graphene sheets for both fundamental and applied
sciences. Therefore, understanding the impact of substrates is crucial
for graphene science and technology.

Charged impurities,\cite{adamtheory,ando} ripples,\cite{ripples} and
resonant scatterers \cite{ostrovsky,stauber,titov,wehling09} have been
considered for modeling the transport property of graphene
field-effect transistors (FETs). Previous experimental studies have
explored the impact of charged impurities \cite{chencharge} and
resonant scatterers \cite{chenhole, rotenberg,geimh, geimresonant} by
using adsorbed impurities or creating vacancies on graphene
sheets. Yet, these studies revealed only the impact of adsorbates or
vacancies and did not shed information on the nature of the native
scatterers already present in the samples. Furthermore, experiments
using different dielectric environments have provided contradictory
results on the role and importance of charged
impurities.\cite{geimdielectric,chaun} Thus, there are no conclusive
experimental results revealing the nature of the native scatterers
that limit the transport properties of graphene on SiO$_2$.

We have measured the impact of low-energy atomic hydrogen on the
transport properties of graphene as a function of coverage and the
initial field-effect mobility. Our transport and Raman spectroscopy
measurements show that hydrogen exerts a short-range scattering
potential which introduces intervalley scattering. Hydrogen transfers
a small but finite amount of charge, as indicated by the
gate-dependent transport measurements. The resistivity added by
hydrogen remains proportional to the number of adsorbed hydrogen and,
therefore, adheres to Matthiessen's rule even at the highest
coverage. This shows that adsorbed hydrogen remains rather dilute and
does not interfere with other pre-existing scattering mechanisms. The
added resistivity at high carrier densities varies approximately as
$n^\delta$, where $n$ is the carrier density and $\delta \approx
-1.5$. Importantly, the saturation coverage of atomic hydrogen is
found to be proportional to the inverse initial mobility and,
therefore, to the number of pre-existing scattering sites. Finally,
our results show that the reactivity to atomic hydrogen is a
characteristic manifestation of the most dominant scatterer in
graphene sheets on SiO$_2$.

The graphene FETs in our measurements are prepared using the
conventional method.\footnote{Graphene is obtained from Kish graphite
  by mechanical exfoliation \cite{novoselovpnas} on 280 nm SiO$_2$
  over doped Si, which is used as the back gate. Raman spectroscopy is
  used to confirm that the samples are single layer
  graphene.\cite{ferrari} Au/Cr electrodes, defined by electron-beam
  lithography, contact the graphene sheets. The devices are annealed
  in H$_2$/Ar at 300 $^o$C for 1 hour to remove resist
  residues.\cite{ishigamiatomic}. Each device is annealed in
  ultra-high vacuum at 490 K for longer than 8 hours to eliminate any
  residual adsorbates prior to hydrogen dosing experiments.} Transport
properties are measured using the four-probe method.\footnote{The
  voltage probes were placed on graphene areas measuring (length
  $\times$ width) 5.0 $\mu$m$\times$6.4$\mu$m (sample A), 0.74
  $\mu$m$\times$0.31$\mu$m (sample B), and 7.8 $\mu$m$\times$8.2
  $\mu$m (sample C).} The initial, pre-hydrogenation mobility ranged
from 1900 to 8300 cm$^2$/V sec for different graphene devices. Each
device is hydrogenated at constant temperature between $11\sim 20$
K.\footnote{Dosing is done at a constant temperature, which varied
  between different devices.} We use a commercial atomic hydrogen
cracker, EFM H from Omicron GmBH, which utilizes a tungsten capillary
heated to 2500 K by an electron beam. The cracker also generates high
energy ions which are steered away from graphene using an electric
deflector. The dosage rate of atomic hydrogen \footnote{The dosage
  rate is estimated from the angular distribution of atomic hydrogen
  provided by the manufacturer.} is maintained constant throughout the
measurements using a variable leak valve. The total dosage or
accumulated areal dose density can be very different from the actual
hydrogen coverage depending on the sticking coefficient. Transport
properties are measured at increasing dosages.

\begin{figure}[ht]
\centering
\includegraphics[width=8.5cm]{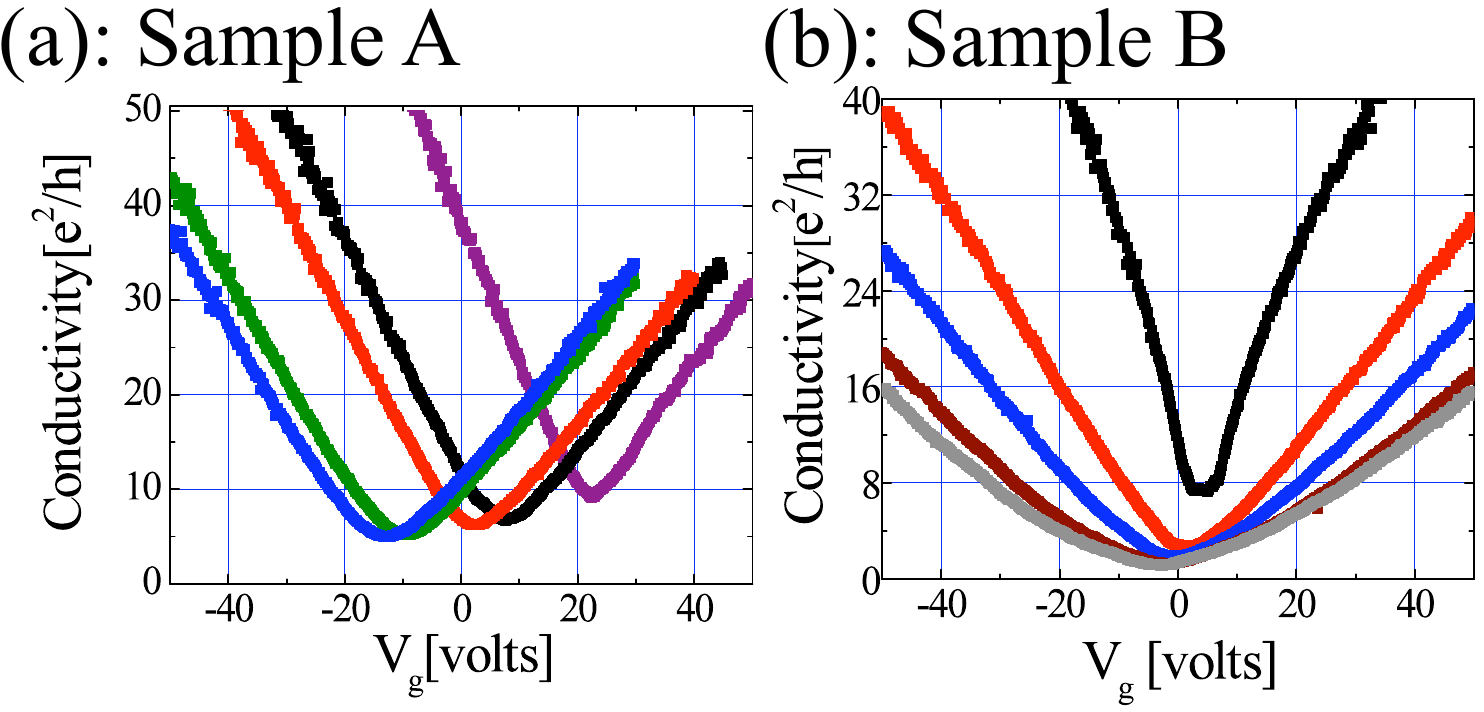}
\caption{(a,b) Impact of atomic hydrogen on the transport properties
  of graphene sheets (samples A and B) for increasing areal dosage
  density. Sample A was measured at 12 K and B at 20 K. The areal
  densities, the number of impinging hydrogen (which may not be
  necessarily adsorbed on graphene), are (a) purple: clean (zero),
  black: $1 \times 10^{15}$/cm$^2$, red: $1.6 \times 10^{15}$/cm$^2$,
  green: $4 \times 10^{15}$/cm$^2$, blue: $5.4 \times 10^{15}$/cm$^2$,
  and (b) black: clean (zero), red: $1.4 \times 10^{14}$/cm$^2$, blue:
  $2.8 \times 10^{14}$/cm$^2$, brown: $5.6 \times 10^{14}$/cm$^2$, and
  silver: $8.5 \times 10^{14}$/cm$^2$.}
\label{fig:1}
\end{figure}

Figure \ref{fig:1} shows the impact of atomic hydrogen adsorption on
the conductivity of graphene sheets. The changes induced by hydrogen
adsorption saturate above a certain dosage. These changes are: (i) a
shift in the gate voltage at which the conductivity is minimal
($V_{\rm min}$), (ii) an increase in the intensity of the D peak in
the Raman spectra, (iii) a monotonic decrease of the conductivity
minimum, and (iv) an additional gate-dependent resistivity which
varies as $|V_{\rm g} - V_{\rm min}|^\delta$ where $\delta \approx
-1.5$ at large $|V_{\rm g} - V_{\rm min}|$. The gate dependence of the
conductivity becomes superlinear at high dosage levels as a result of
this exponent. Below, we discuss each change in more detail.

A finite charge is donated to graphene by the adsorbed hydrogen, as
indicated by the shift of $V_{\rm min}$ upon hydrogenation. The
observed sign of the charge transfer from atomic hydrogen to carbon is
consistent with a previous experiment \cite{heinz} and theoretical
calculations,\cite{zhu, carneiro} but different from hydrogenation
studies using atomic hydrogen derived from a hydrogen
plasma.\cite{geimh,geimresonant} It is not possible to determine the
amount of charge transferred per adsorbed hydrogen directly from our
experiment, as the sticking coefficient of hydrogen on graphene is
unknown. Previous experiments \cite{heinz,geimh,geimresonant} do not
agree on the amount of charge transfer from hydrogen. Theoretical
studies show 0.076 to 0.161 $e$ $ $\cite{carneiro} or 0.16 to 0.25 $e$
$ $\cite{zhu} donated per hydrogen ($e$ denotes the electron charge),
depending on the degree of allowed lattice relaxation \cite{carneiro}
or the position of hydrogen.\cite{zhu} Below, these calculated values
are used to estimate the saturation coverage of hydrogen.

\begin{figure}[ht]
\centering
\includegraphics[width=8.5cm]{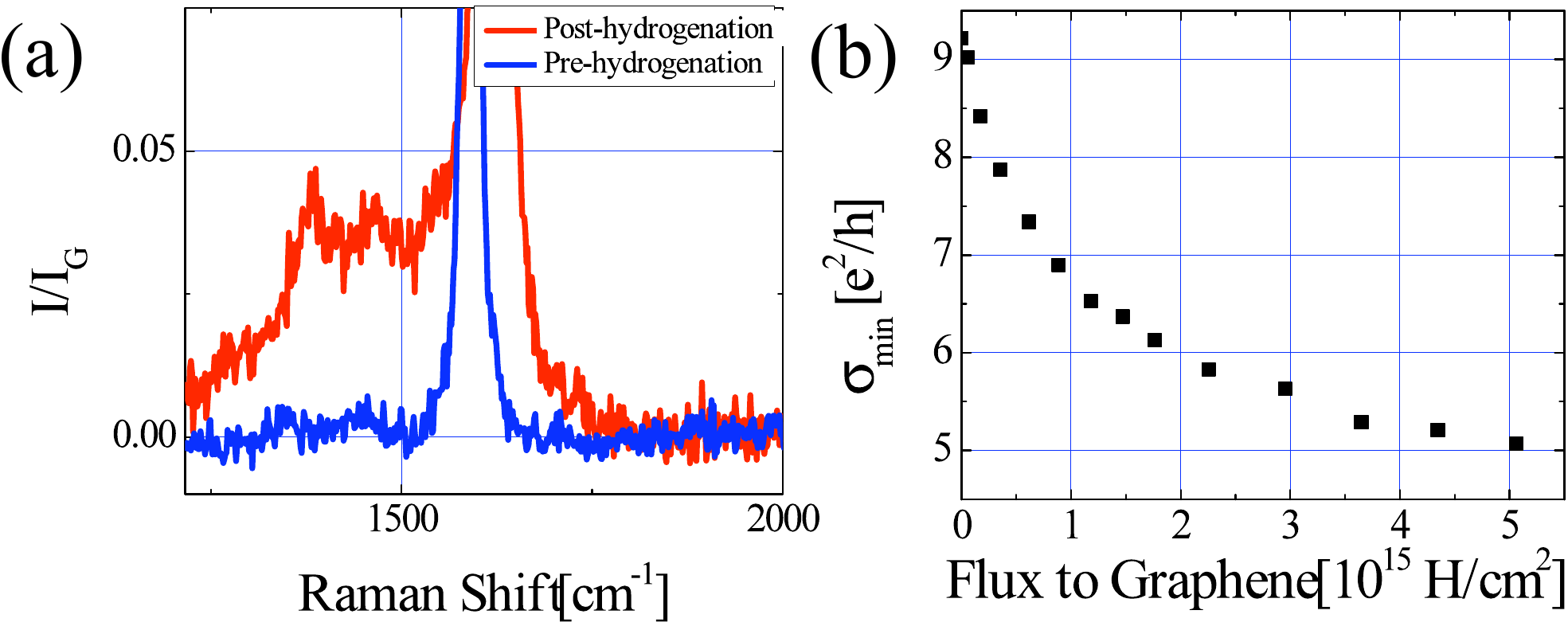}
\caption{(a) Raman spectra acquired for sample A before and after
  hydrogenation. The observed intensity has been normalized to the
  peak height of the G-band. (b) Minimum conductivity as a function of
  increasing dosage in sample A.}
\label{fig:2}
\end{figure}

Raman spectroscopy and the impact of atomic hydrogen on the minimum
conductivity reveal that atomic hydrogen introduces intervalley
scattering and, therefore, exerts a short-range scattering
potential.\cite{ramanreview} Figure \ref{fig:2}a shows Raman spectra
acquired at room temperature in air both before hydrogen dosing and
after achieving saturation at low temperature. The intensity of the D
peak in the Raman spectrum is larger upon adsorbing hydrogen. The
relative intensity of the D peak to the G peak, $I_D/I_G$, which can
be used to estimate the adsorbed hydrogen density,\cite{lucchese, cancado, dband} is 0.0034$\pm$0.0021 and
0.0182$\pm$0.0056 before and after hydrogen adsorption,
respectively.\footnote{Multiple Lorentzian peaks near 1250, 1350, and
  1450 cm${-1}$ were fitted to the experimental data to determine the
  peak height.} The small values observed for this D-G ratio even at
saturation are likely due to the small 
desorption barrier for hydrogen, as discussed below. Figure
\ref{fig:2}b shows that the minimum conductivity decreases
monotonically as a function of hydrogen dosage. The minimum
conductivity at saturation ranges from 0.52 to 5.1 $e^2/h$ for
different devices. Since long-range scatterers have been found to vary
the minimum conductivity non-monotonically and not below 4
$e^2/h$,\cite{adamtheory,chencharge} our transport measurements are
also consistent with hydrogen exerting a short-range scattering
potential.

\begin{figure}[ht]
\centering
\includegraphics[width=8.5cm]{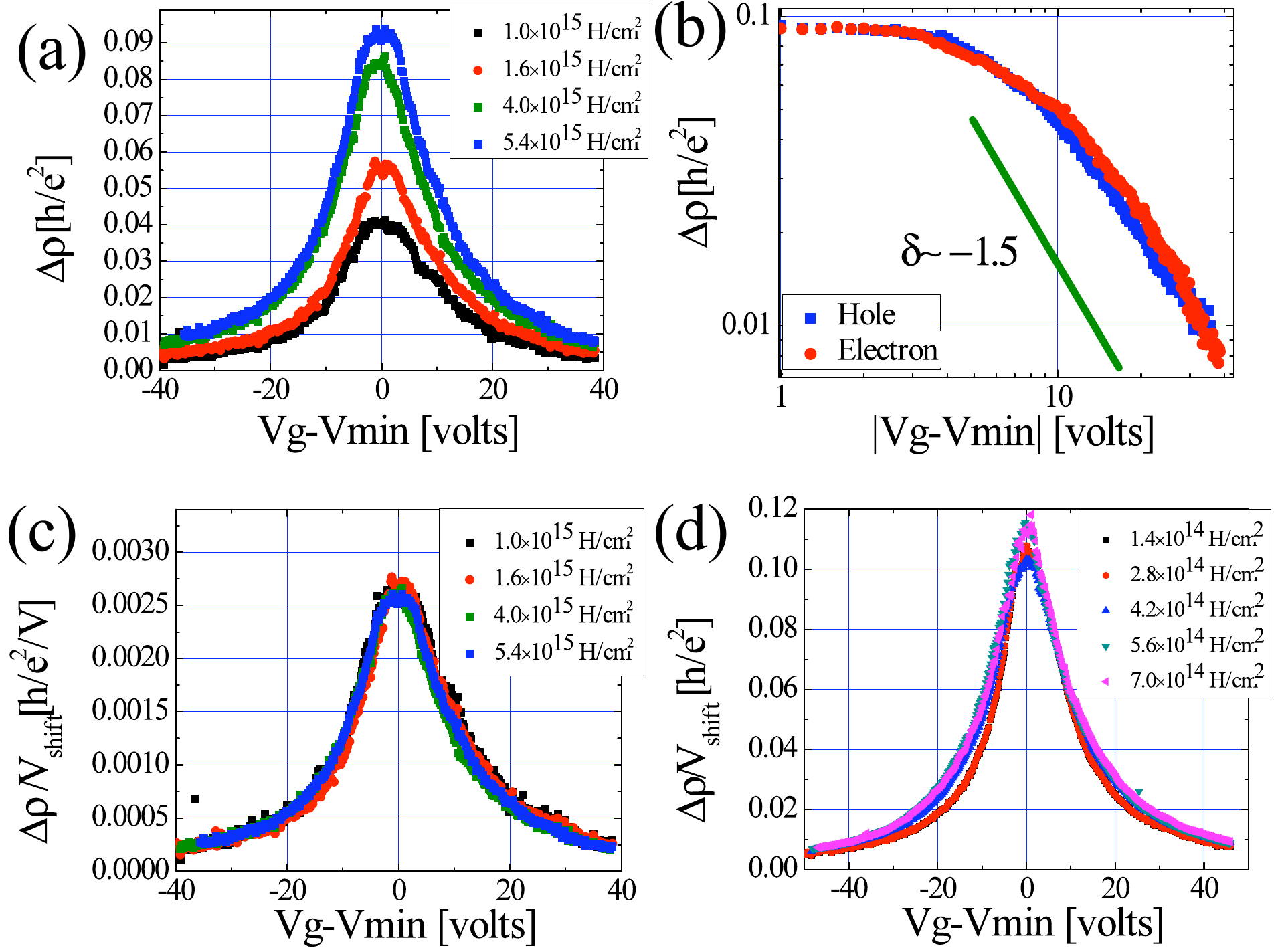}
\caption{(a) Resistivity added by hydrogen as a function of
  $V_g-V_{\rm min}$ at different areal dosage density. (b) Gate
  dependence of the added resistivity as a function of $V_g-V_{\rm
    min}$ at the areal dosage density of $5.4 \times 10^{15}$
  H/cm$^2$. The green line indicates the slope for an exponent of
  -1.5. (c) Added resistivity as a function of $V_g-V_{\rm min}$ at
  different areal dosage normalized to $V_{\rm shift}$. [(a)-(c) for
    sample A] (d) Same as in (c), but for sample B.}
\label{fig:3}
\end{figure}

Figure \ref{fig:3}a shows the added resistivity due to atomic hydrogen
at different dosage levels as a function of $V_{\rm g}-V_{\rm
  min}$. The impact of atomic hydrogen is nearly electron-hole
symmetric and the added resistivity varies approximately as $|V_{\rm
  g}-V_{\rm min}|^{-1.5}$ at large $|V_{\rm g} - V_{\rm min}|$ for all
samples, as shown in Fig. \ref{fig:3}b. The resistivity exponent
differs from the -1 value expected for Coulomb impurities
\cite{adamtheory,chencharge} and the electron-hole symmetry is
consistent with a resonant scatterer positioned very close to the
Fermi level (i.e. a mid-gap resonant state).\cite{peres} The observed
exponent also agrees with calculated exponents for resonant scatterers
with a finite on-site amplitude \cite{muccioloperes} as well as for
Gaussian-correlated scatterers.\cite{adamxover}$^,$\footnote{Previous
  calculations \cite{stauber,titov,wehling10} using resonant
  scatterers with infinite on-site energy yield a resistivity exponent
  less or equal to -1.} As shown in Fig. \ref{fig:3}c, we find that
the curves of added resistivity versus gate voltage for successive
dosage levels collapse on top of each other when divided by the
induced shift in $V_{\rm min}$, $V_{\rm shift}$, indicating that the
added resistivity at different dosage levels is proportional to
$V_{\rm shift}$. Therefore, the number of adsorbed hydrogen atoms is
directly proportional to $V_{\rm shift}$. For long-range scatterers
such as potassium adsorbates,\cite{chencharge} $V_{\rm shift}$ does
not vary linearly with the number of adsorbates. Such non-linearity
has been attributed to incomplete screening of the potential imposed
by potassium on graphene.\cite{adamtheory} Therefore, we conclude that
the excess charge of adsorbed atomic hydrogen is effectively screened
by graphene. All samples we have measured show a similar behavior (for
instance, see Fig. \ref{fig:3}d for sample B). Deviations from the observed
normalization by $V_{\rm shift}$ are found only at low
dosages and can be attributed to uncertainty in
determining $V_{\rm min}$ at low dosage. The observed normalization
also shows that the scattering cross section of hydrogen does not vary
appreciably even at higher dosage levels and that hydrogen does not
modify other scattering mechanisms. Therefore, the added resistivity
by hydrogen follows Matthiessen's rule, $R_{\rm total} = R_{\rm
  adsorbates} + R_{\rm substrate} + R_{\rm graphene}$, where $R_{\rm
  adsorbates}$ and $R_{\rm substrate}$ are due to scattering by
adsorbates and the substrate, respectively, and $R_{\rm graphene}$ is
the intrinsic resistance of the graphene sheet.

\begin{figure}[ht]
\centering
\includegraphics[width=8.5cm]{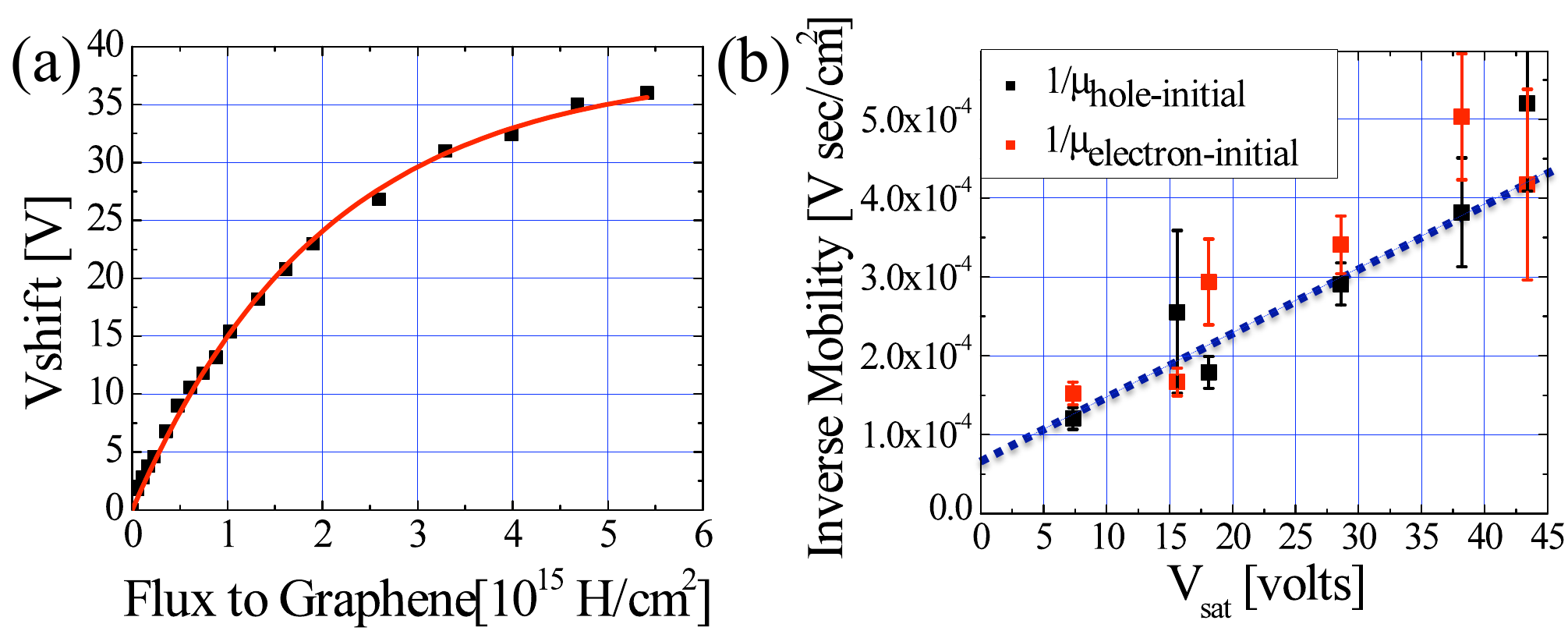}
\caption{(a) $V_{\rm shift}$ as a function of the increasing areal
  dosage density for sample A. (b) Initial maximum electron and hole
  mobility as a function of the saturation voltage shift, $V_{\rm
    sat}$, for different samples.}
\label{fig:4}
\end{figure}

Figure \ref{fig:4}a shows $V_{\rm shift}$, which is proportional to
the number of adsorbed hydrogen, as a function of the accumulated
hydrogen dosage. The behavior is well described by a saturating
exponential function, with a saturation voltage denoted $V_{\rm
  sat}$. A wide range of $V_{\rm sat}$ is observed for different
samples, from 7.34 to 43.4 volts. The maximum shift of 43.4 volts
implies that the observed maximum coverage of hydrogen is 0.012
assuming the predicted charge transfer \cite{carneiro} of 0.076 $e$
per adsorbed hydrogen. We find no correlation between experimental
temperatures and saturation voltages.
Figure \ref{fig:4}b shows that the saturation coverage for different
samples is inversely proportional to their initial maximum electron and hole
field-effect mobility. \footnote{$\mu=\frac{1}{c_g}\frac{d \sigma}{dV_g}$, where $\mu$ is field effect mobility and $c_g$ is capacitance per unit area, was used to calculate the gate-dependent mobility.} Since the inverse mobility is proportional to
the number of scatterers, our data show that the number of native
scatterers is proportional to the number of possible adsorption sites
for hydrogen. By extrapolation to the limit where these sites are
absent, we obtain a mobility of $(1.5 \pm 0.3) \times 10^4$ cm$^2$/V
sec, as determined by a linear fit. This extrapolated mobility value
is still an order of magnitude lower than the field-effect mobility
measured for suspended graphene sheets,\cite{bolotinss,bolotinprl}
showing that there are still other, less important, scatterers
reducing the mobility of graphene on SiO$_2$. Interestingly, the
extrapolated value is similar to the maximum mobility observed on
SiO$_2$ in previous studies \cite{geimmobility,tan} suggesting that
the reactivity to hydrogen is the signature of the most dominant type
of native scatterers for all graphene devices on SiO$_2$.

\begin{figure}[ht]
\centering
\includegraphics[width=5cm]{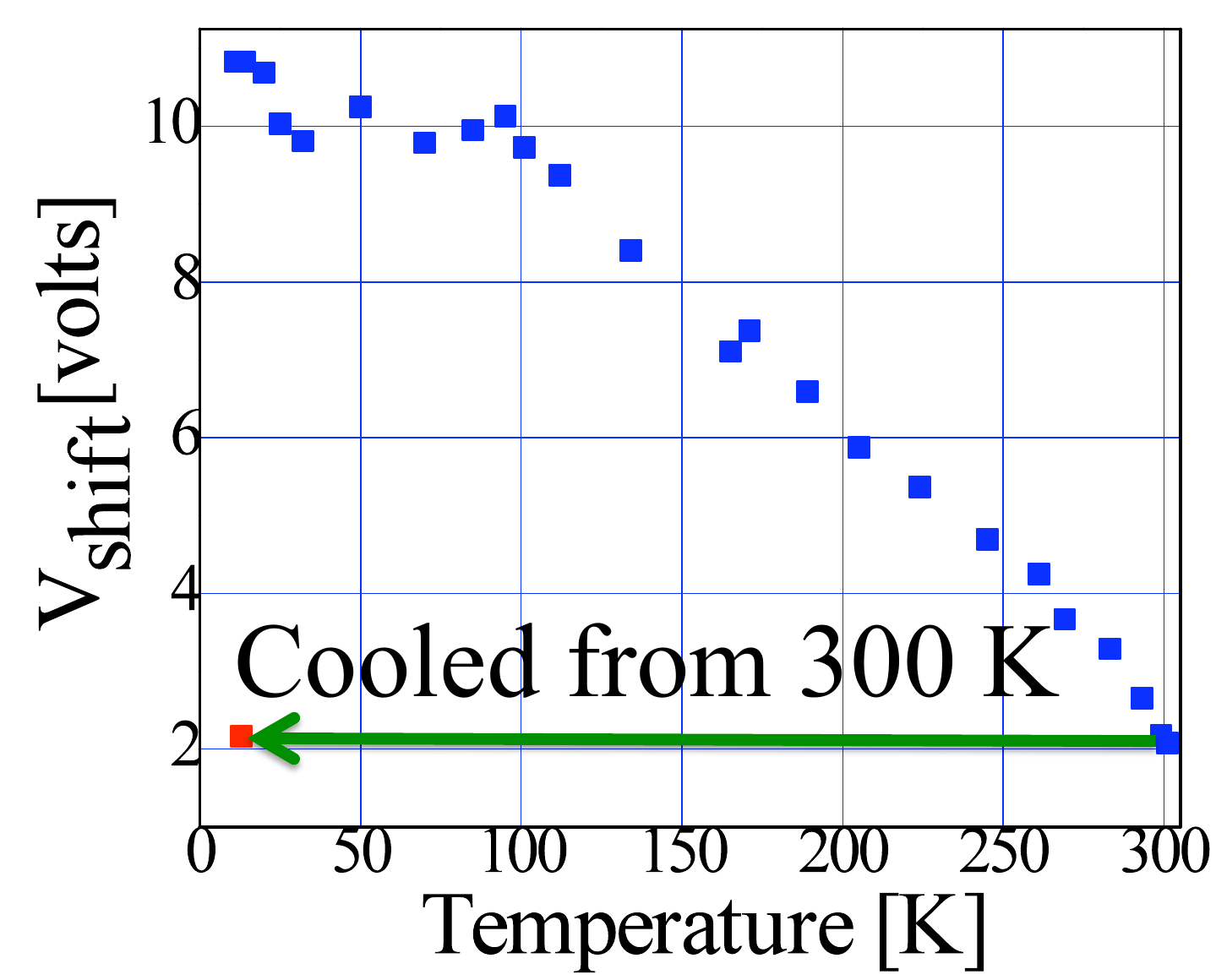}
\caption{$V_{\rm shift}$ at increasing temperature for sample C after
  reaching saturation coverage by atomic hydrogen at 11 K. Data
  acquired at a warming rate of 0.45 to 6 K/min. Red point indicates
  $V_{\rm shift}$ when the warmed hydrogenated device is cooled down
  again from 300 K.}
\label{fig:5}
\end{figure}

$V_{\rm shift}$ induced by the adsorbed hydrogen is reduced as the
temperature is raised, as shown in Fig. \ref{fig:5}. The value of
$V_{\rm shift}$ remains constant when warmed samples
are again cooled, indicating that the observed
reduction in $V_{\rm shift}$ is due to the desorption of
hydrogen. This temperature dependence indicates that the desorption
energy of adsorbed hydrogen on graphene is much smaller than the
previously reported values of approximately 1 eV on
graphite.\cite{zecho} A small desorption energy explains the small D
peak observed in the Raman spectra of the hydrogen-dosed samples
acquired at room temperature and suggests that atomic hydrogen is not
forming a fully relaxed covalent bond to carbon. Furthermore, we also
know that the maximum thermal energy of impinging atomic hydrogen
barely exceeds the barrier of 0.21 eV calculated for the attachment of
atomic hydrogen to planer graphite.\cite{Jackson} Therefore, atomic
hydrogen is binding only to unusual, chemically-activated sites, which
do not relax to a full sp$^3$ configuration upon adsorbing hydrogen.

It is possible that the reactivity of graphene sheets is enhanced by
adding curvature or changing the Fermi level. Wrinkles
\cite{guinea,pereira} and ripples\cite{ripples} can perturb the sp$^2$
bonds, generating chemically-activated sites for hydrogen. Charge
puddles \cite{adamtheory} may also increase the reactivity of graphene
sheets. However, the data presented in this paper cannot determine the sites with affinity to atomic hydrogen in graphene on
SiO$_2$. Atomically-resolved microscopy on hydrogenated graphene
devices can reveal the exact mechanism for the enhancement of the
reactivity, correlated to the most dominant scatterer in graphene on
SiO$_2$. These studies are being presently carried out.

In conclusion, we used atomic hydrogen to probe the nature of native
scatterers in graphene. Hydrogen exerts short-range scattering
potential in graphene, as indicated by Raman spectroscopy and the
impact on the minimum conductivity. Charge is transferred from
hydrogen to carbon and the Coulomb potential created by the induced
charge on hydrogen is effectively screened by carriers in
graphene. The adherence of the added resistivity to Matthiessen's rule
also shows that: (1) adsorbates do not influence the resistivity
caused by other factors (such as lattice defects, phonons, etc.) and
(2) the number of adsorbed hydrogen, $n_H$, is proportional to $V_{\rm
  shift}$. Finally, the number of hydrogen adsorption sites is found
to correspond to the number of native scatterers; in the absence of
these scatterers, the carrier mobility of graphene sheets will reach
$1.5\times 10^4$ cm$^2$/V sec. The scatterers uncovered in this study
dominate the transport properties of graphene-based FETs on SiO$_2$ and the
affinity to atomic hydrogen is the hallmark of these scatterers.
Our results provide an important insight into the nature of the scatterers which limit mobility of graphene sheets on substrates.

This work is based upon research supported by the National Science
Foundation under Grant No. 0955625. MI thanks Enrique del Barco and
Simranjeet Singh for collaboration in developing the graphene device
fabrication procedure at UCF and providing access to photolithography
and electron beam evaporator facilities. EM thanks Rodrigo Capaz,
Vitor Pereira, and Nuno Peres for fruitful discussions.

\bibliography{references_em}

\end{document}